\documentclass[12pt,reqno]{amsart}
\usepackage{mathrsfs}
\usepackage{amsfonts}
\allowdisplaybreaks[4]
\usepackage{graphicx} 
\usepackage{epsfig}
\usepackage{amssymb}
\usepackage{epsfig}
\usepackage{amsmath,amsfonts}
\usepackage{cite}
\usepackage{subfigure}

\theoremstyle{plain}
\newtheorem{theorem}{Theorem}[section]

\theoremstyle{definition}

\newtheorem{remark}[theorem]{Remark}

\numberwithin{equation}{section}
\begin{document}
\title[Solitons ]{High-order
rogue waves for the Hirota equation}
\author{Linjing Li, Zhiwei Wu, Lihong Wang, Jingsong He$^*$}
\thanks{$^*$Corresonding Author: Email: hejingsong@nbu.edu.cn, Tel: 86-574-87600739, Fax: 86-574-87600744}
\dedicatory { \dag \ Department of Mathematics, Ningbo University,
Ningbo, Zhejiang 315211, P.\ R.\ China }

\begin{abstract}
The Hirota equation is better than the nonlinear Schr\"{o}dinger equation
when approximating deep ocean waves. In this paper, high-order rational solutions for the Hirota equation are constructed based on the parameterized Darboux transformation. Several types of this kind of solutions are classified by their structures.
\end{abstract}

 \maketitle \vspace{-0.6cm}


\noindent {{\bf PACS number(s)}: 42.65.Tg, 47.35.Fg, 05.45Yv

\section{Introduction}
In recent decades, the rogue waves\cite{muller}(or freak
waves\cite{draper}), have gained more and more interests in the
study of physics. One of the reason is that such kind of waves emerges
spontaneously and frequently in the ocean without any
sign, ``appears from nowhere and disappears without a
trace\cite{akhmediev}". Recently, besides oceanography, the rogue waves are also found in
the optical systems when
generating highly energetic optical pulses\cite{hasegawa1,anderson}, as well as in the superfluids and femtosecond
pulse\cite{nnakhmediev,hasegawa2,tai1,tai2,potasek}, etc. The
growing consensus, which is from the oceanic and optical rogue
waves, arises from to the modulation
instability(MI)\cite{garrett,onorato,dudley,vysloukh,akhmediev4} which is a key property for many integrable equations. One of the typical
example is the nonlinear Schr\"{o}dinger(NLS) equation

 \begin{equation}\label{nls}
  iq_t+2|q|^{2}q+q_{xx}=0.
 \end{equation}
For this equation, the lower-order rogue wave has been
obtained\cite{Peregrine} with a good experimental simulation. People
are more interested in the higher order waves, because its numerous
and complicated properties. Some special higher order rogue waves
have been constructed by Darboux transformation\cite{akhmediev1,he},
and some more general cases (multi-Peregrine) were well studied
\cite{dubard,matveev,gaillard,akhmediev2,jianke}. Unlike soliton
solutions siting on the zero background, these higher order rogue
waves not only sit on the non-zero constant background, but lots of
behaviors are quite different at points on the space-time
plane\cite{akhmediev2}. On the other hand, the rogue waves are the
limit of a sequence of the breather solutions\cite{clarke,he}. These
new studies about rogue waves enrich the conception of rogue waves
and lead to understand further towards this mysterious phenomenon.

However, the phenomena of the ocean are significantly complicated,
such as the depth of the sea, bottom friction, viscosity, etc, these
factors all need to be taken into account when we do the simulation.
Therefore we need to add higher order terms and other nonlinear
effects\cite{ohkuma,christo}. The Hirota equation can be viewed as a
generalization for the NLS equation, when the higher dispersion and
time-delay changes is taken into account.

 In dimensionless form, the Hirota equation is given
by
\begin{equation}\label{hirota}
iq_t+\alpha(2|q|^{2}q+q_{xx})+i\beta(q_{xxx}+6|q|^{2}q_x)=0, \alpha, \beta\in\mathbb{R}.
\end{equation}
The variable $t$ is the propagation variable and $x$ is the
retarded time variable in a moving frame. The phase function $q$ is the envelope of
the wave field. The two terms in
(\ref{hirota}) that enter with a real coefficient $\beta$
represent the third-order dispersion ($q_{xxx}$) and a time-delay
correction to the cubic term ($|q|^{2}q_x$), respectively. We can see
that
\begin{enumerate}
\item $\alpha=1,\beta=0$, (\ref{hirota}) becomes the NLS
equation: $iq_t+2|q|^{2}q+q_{xx}=0,$
\item $\alpha=0,\beta=1$, (\ref{hirota}) becomes the modified
KdV (mKdV) equation:\\
 $q_t+q_{xxx}+6|q|^{2}q_x=0.$
\end{enumerate}

In some cases, if an integrable equation is a combination of two
known equations that are also integrable, the low-order solutions
can be derived from solutions of those two equations. But for
high-order rogue waves, it does not work. Therefore, we provide the
parameterized Darboux transformation to derive the high-order rational solutions.

The organization of this paper is as follows. In section 2, based on
the parameterized Darboux transformation, the general formation of the solution is given in terms of the Gram determinants. In section 3, we construct the higher order rogue waves from a periodic seed with a constant amplitude and make a classification based on their structures. We provide further discussion in section 4.

\section{The Darboux Transformation for the Hirota equation}
In this section, we give the explicit Darboux transformation for the
Hirota equation.

The Lax pair for the Hirota equation is as follows\cite{taohe}:
\begin{equation}\label{Lax pair}
\begin{cases}
\psi_x&=M\psi, \\
\psi_t&=N\psi,
\end{cases}
\end{equation}
where
\begin{align}
 &\psi=\left(
\begin{array}{c}
\phi_1\\\phi_2
\end{array}\right),
M=\left(
\begin{array}{cc}
-i\lambda & q\\
r & i\lambda
\end{array}\right), N=V_3\lambda^{3}+V_2\lambda^{2}+V_1\lambda+V_0,\\
&V_3=\left(
\begin{array}{cc}
-4i\beta & 0\\0 & 4i\beta
\end{array}\right),
V_2=\left(
\begin{array}{cc}
-2i\alpha & 4q\beta\\4\beta\gamma & 2i\alpha
\end{array}\right),
V_1=\left(
\begin{array}{cc}
-2iqr\beta & 2iq_x\beta+2q\alpha\\-2ir_x\beta+2r\alpha & 2iqr\beta
\end{array}\right), \\
&V_0=\left(
\begin{array}{cc}
-iqr\alpha+\beta(-qr_x+q_xr) &
iq_x\alpha-\beta(q_{xx}-2q^{2}r)\\-ir_x\alpha+\beta(-r_{xx}+2qr^{2})
& iqr\alpha-\beta(-qr_x+q_xr)
\end{array}\right).
\end{align}
Here $\lambda$ is a spectral parameter with complex value, and
$\psi$ is the eigenfunction corresponding to the $\lambda$ of the
Hirota equation.

In terms of zero-curve equation
$M_t-N_x+[M,N]=0$\label{zero}, equations for $q$ and $r$ are as follows:
\begin{equation}\label{qr}
\begin{cases}
q_t=i\alpha(2|q|^{2}q+q_{xx})-\beta(q_{xxx}+6|q|^{2}q_x),\\
r_t=-i\alpha(2|r|^{2}r+r_{xx})-\beta(r_{xxx}+6|r|^{2}r_x),
\end{cases}
\end{equation}
These two equations admit the constraint
\begin{equation}\label{constrain}
r=-q^{*},
\end{equation}
 here $*$ denotes the complex conjugate. After the constraint, the system (\ref{qr}) becomes one equation, which is exactly the Hirota equation.

 Next, we will construct a gauge transformation
\begin{equation}\label{gauge}
\psi^{[1]}=T\psi,
\end{equation}
such that, after this gauge transformation,
\begin{equation}\label{xt}
\begin{cases}
\psi^{[1]}_{x}=M^{[1]}\psi^{[1]},\ M^{[1]}=(T_x+TM)T^{-1}.\\
\psi^{[1]}_{t}=N^{[1]}\psi^{[1]},\ N^{[1]}=(T_t+TN)T^{-1}.\\
\end{cases}
\end{equation}
is solvable for $\psi^{[1]}$ and $M^{[1]}$, $N^{[1]}$ are of the same form as $M$,
$N$. Moreover, we want $M^{[1]}$, $N^{[1]}$ also
admits the constraint (\ref{constrain}). Therefore, we can get new
solution of the Hirota equation from one solution.

Let $\{\psi_j; 1\leq j \leq n \}$ be eigenfunctions for the Lax
pair (\ref{Lax pair}) which are corresponding to eigenvalues
$\lambda_{j}, j=1,2,...,n$:
 \begin{equation}
\psi_j=\left(
\begin{array}{cc}
\phi_{j1}(x,t,\lambda_j)\\ \phi_{j2}(x,t,\lambda_j)
\end{array}\right),\ j=1,2,......n.
\end{equation}
Moreover, if $\psi_j$ is a eigenfunction for eigenvalue $\lambda_j$,
then
\begin{equation*}
\psi'_j:=\left(
\begin{array}{cc}
-\phi^{*}_{j2}\\ \phi^{*}_{j1}
\end{array}\right)
\end{equation*}
is the eigenfunction corresponding to $\lambda^{*}_{j}$.

Choose a sequence of eigenvalues \{ $\lambda_{j}|
j=1,2,3,...,2n$\}, such that $\lambda^{*}_{2n-1}=\lambda_{2n}$ and
\{$\psi_{1},\psi_{2},\psi_{3},...,\psi_{2n}$\} is the corresponding
eigenfunctions. We use a modified mechanism described in \cite{he} to construct the parameterized Darboux transformation for the Hirota equation:

\noindent {\bf Theorem 1.}{\sl Given $\lambda_{1}\in\mathbb{C}$, and
\begin{equation*}
 \psi_{1}=\left(
\begin{array}{cc}
\phi_{11}\\ \phi_{12}
\end{array}\right)
\end{equation*} is an eigenfunction corresponding to $\lambda_{1}$ of
(\ref{Lax pair}), denote
\begin{eqnarray}\label{T1}
T_1(\lambda;\lambda_1)=\left(
\begin{array}{cc}
1 & 0\\0 & 1
\end{array}\right)\lambda+
\left(
\begin{array}{cc}
-\frac{1}{\Delta_2}\left|\begin{array}{cc}
 \lambda_1\phi_{11}&\phi_{12} \\
\lambda_2\phi_{21}&\phi_{22}
 \end{array}\right|&
\frac{1}{\Delta_2}\left|\begin{array}{cc}
 \lambda_1\phi_{11}&\phi_{11} \\
\lambda_2\phi_{21}&\phi_{21}
 \end{array}\right|\\
\frac{1}{\Delta_2}\left|\begin{array}{cc}
 \phi_{12}&\lambda_1\phi_{12} \\
\phi_{22}&\lambda_2\phi_{22}
 \end{array}\right| &-\frac{1}{\Delta_2}\left|\begin{array}{cc}
 \phi_{11}&\lambda_1\phi_{12} \\
\phi_{21}&\lambda_2\phi_{22}
 \end{array}\right|  \\
\end{array}  \right),\label{DT1matrix}
\end{eqnarray}
with $\Delta_2=\left|\begin{array}{cc}
 \phi_{11}&\phi_{12} \\
\phi_{21}&\phi_{22}
 \end{array}\right|$,
then
\begin{equation}
\psi^{[1]}=T_1\psi_1
\end{equation}
is again a solution of (\ref{Lax pair}) and
\begin{eqnarray}\label{t1q}
q_{1}=q+2i\frac{1}{\Delta_2}\left|\begin{array}{cc}
 \lambda_1\phi_{11}&\phi_{11} \\
\lambda_2\phi_{21}&\phi_{21}
 \end{array}\right|,
\end{eqnarray}
is a new solution for the Hirota equation.}

This formula is of the same form as in \cite{he}. For high-order rational solutions $q_n, r_n$, calculation is quite cumbersome, but one can still check these equations with the aid of modern tools such as ``maple" or equivalent, and also numerically.

\noindent {\bf Theorem 2 (N-fold).} {\sl By applying the gauge
transformation repeatedly for N times, choose eigenvalues
\{$\lambda_{1},\lambda_{2},\lambda_{3},...,\lambda_{2n}$ \}, such
that $\lambda^{*}_{2n-1}=\lambda_{2n}, n=1,2,...,N$, and $\psi_{j}$
is an eigenfunction for $\lambda_{j}$. Then we can get N-fold
solution for the Hirota equation:
\begin{eqnarray}\label{nT}
q_{n}=q-2i\frac{A_{2n}}{B_{2n}},
\end{eqnarray}
where
\begin{equation}\label{numer}
A_{2n}=\begin{vmatrix}
\phi_{11}&\phi_{12}&\lambda_1\phi_{11}&\lambda_{1}\phi_{12}&\ldots&\lambda_{1}^{n-1}\phi_{11}&\lambda_{1}^{n}\phi_{11}\\
\phi_{21}&\phi_{22}&\lambda_2\phi_{21}&\lambda_{2}\phi_{22}&\ldots&\lambda_{2}^{n-1}\phi_{21}&\lambda_{2}^{n}\phi_{21}\\
\phi_{31}&\phi_{32}&\lambda_3\phi_{31}&\lambda_{3}\phi_{32}&\ldots&\lambda_{3}^{n-1}\phi_{31}&\lambda_{3}^{n}\phi_{31}\\
\phi_{41}&\phi_{42}&\lambda_4\phi_{41}&\lambda_{4}\phi_{42}&\ldots&\lambda_{4}^{n-1}\phi_{41}&\lambda_{4}^{n}\phi_{41}\\
\vdots&\vdots&\vdots&\vdots&\vdots&\vdots&\vdots\\
\phi_{2n1}&\phi_{2n2}&\lambda_{2n}\phi_{2n1}&\lambda_{2n}\phi_{2n2}&\ldots&\lambda_{2n}^{n-1}\phi_{2n1}&\lambda_{2n}^{n}\phi_{2n1}\nonumber\\
\end{vmatrix},
\end{equation}

\begin{equation}\label{denom}
B_{2n}=\begin{vmatrix}
\phi_{11}&\phi_{12}&\lambda_1\phi_{11}&\lambda_{1}\phi_{12}&\ldots&\lambda_{1}^{n-1}\phi_{11}&\lambda_{1}^{n-1}\phi_{12}\\
\phi_{21}&\phi_{22}&\lambda_2\phi_{21}&\lambda_{2}\phi_{22}&\ldots&\lambda_{2}^{n-1}\phi_{21}&\lambda_{2}^{n-1}\phi_{22}\\
\phi_{31}&\phi_{32}&\lambda_3\phi_{31}&\lambda_{3}\phi_{32}&\ldots&\lambda_{3}^{n-1}\phi_{31}&\lambda_{3}^{n-1}\phi_{32}\\
\phi_{41}&\phi_{42}&\lambda_4\phi_{41}&\lambda_{4}\phi_{42}&\ldots&\lambda_{4}^{n-1}\phi_{41}&\lambda_{4}^{n-1}\phi_{42}\\
\vdots&\vdots&\vdots&\vdots&\vdots&\vdots&\vdots\\
\phi_{2n1}&\phi_{2n2}&\lambda_{2n}\phi_{2n1}&\lambda_{2n}\phi_{2n2}&\ldots&\lambda_{2n}^{n-1}\phi_{2n1}&\lambda_{2n}^{n-1}\phi_{2n2}\nonumber\\
\end{vmatrix}.
\end{equation}
}

As a consequence, high-order rogue waves for the NLS equation and
the mKdV equation can be derived.

\section{The High-order rogue waves}
In this section, we give high-order rogue waves for the Hirota equation from
non-zero seeds. By using the principle of the superposition of the
linear differential equation, we can get the new eigenfunction
corresponding to $\lambda_j$:
\begin{equation*}
 \psi_{j}=\left(
\begin{array}{cc}
\varphi_{j1}\\ \varphi_{j2}
\end{array}\right)
\end{equation*}
\begin{equation}\label{varphi}
\varphi_{j1}=d_1\phi_{j1}+d_2\phi_{j2},\
\varphi_{j2}=-d_1\phi^{*}_{j2}+d_2\phi^{*}_{j1},
\end{equation}
with
\begin{equation}
d_1=e^{ic_1(s_0+s_1\varepsilon+s_2\varepsilon^{2}+\ldots+s_{n-1}\varepsilon^{n-1})},\
d_2=e^{-ic_1(s_0+s_1\varepsilon+s_2\varepsilon^{2}+\ldots+s_{n-1}\varepsilon^{n-1})},
\nonumber
\end{equation}
\begin{equation}
\phi_{11}=ce^{i[(\frac{1}{2}a+c_1)x+(\frac{1}{2}b+2c_1c_2)t]},\
\phi_{12}=i(\frac{1}{2}a+\lambda_j+c_1)e^{i[(-\frac{1}{2}a+c_1)x
+(-\frac{1}{2}b+2c_1c_2)t]},\nonumber
\end{equation}
\begin{equation}
c_1=\frac{1}{2}\sqrt{a^2+4c^2+4\lambda_ja+4\lambda_j^2},\ \
c_2=\alpha\lambda_j+2\beta\lambda^{2}_j-\frac{1}{2}a\alpha-\beta~c^2+\frac{1}{2}\beta~a^2
-\lambda_j\alpha\beta. \nonumber
\end{equation}
Here $a,b,c, s_i, i=0,1,2,\dots$ are the constants, $\varepsilon$ is
infinitesimal, $\psi_1=(\phi_{11}, \phi_{12})^{T}$ is the solution of (\ref{Lax pair}) for the non-zero seed $q=ce^{i\rho},\ \rho=ax+bt,\
b=\alpha(2c^2-a^2)+\beta(a^3-6ac^2)$.

It is well known that high-order rogue waves can be obtained by the Taylor expansion of the breather solutions.
 In \cite{he}, the scheme for nonlinear schr\"{o}dinger equation is established.
 Due to the complication and tediousness of the explicit formula, we describe the high-order rogue waves by their structures,
explicit formula for solution $q_1$ and $q_2$ will be given in the
Appendix. Moreover, in the rest of this paper, we will dedicate to
the classification of rational solutions.


 In our classification, we can see that by choosing different values for parameters $d_1$ and $d_2$ different types of solutions are obtained.

\begin{enumerate}
\item When $d_1=d_2=1$(after expansion), rational solutions of any order $n$ have similar structure.
\begin{enumerate}
\item There are $\frac{n(n+1)}{2}-1$ local maximum in each side of the line $t=0$.

\item Starting from $-\infty$, before the central high amplitude (the global maximum), there is a sequence of peaks with increasing height. In detail, the number of first peaks is $n$, then there is a row of $n-1$ peaks and so on.

\item  The structure is symmetric with respect to time $t$.
\end{enumerate}
We called this type of rogue wave the $\emph{fundamental pattern}$
(same as in \cite{he} Figure 1).

\item $\emph {Ring}$ structure.

In (\ref{varphi}), when $n\geq3$ and the $\emph {principle coefficient}$ for order $n$ rational solution $s_{n-1}\gg1$,
 while other coefficients $s_{i}$ are all zero. The structure for rational solutions consists of two parts.
  The outer circular shell consists of $2n-1$ first order rational solutions, while the center is an order
   $(n-2)$ rational solution of fundamental pattern. This structure is similar to the ``wave clusters" in \cite{david}. Figure 2 give a visual description for the ring structure.

\item $\emph {Triangular}$ structure.

Another typical type of rational solutions can be obtained by choosing the first nontrivial coefficient $s_{1}\gg1$, while others are all zero. The order $n$ rational solutions have a structure of equilateral triangular type with $n$ peaks in each edge. Each single peak is of the same height and the total number of peaks is $\frac{n(n+1)}{2}$ (Figure 3).
\end{enumerate}

\begin{remark}
When $n=2$, the ring structure and triangular
structure have the same shape: there is a single triangular with
$3=3n-3$ peaks in the triangular structure, and in the ring
structure, since $n-2=0$, there is no basic mode in the middle and
the outer ring consists of $3=2n-1$ peaks.
\end{remark}

 These three types of structures listed above are the basic scheme for high-order rogue waves. Another interesting fact worth emphasized here is that: high-order rogue waves are usually the combination of these three types: `fundamental pattern", ``$\emph {ring}$" structure and ``$\emph {triangular}"$ structure. In the rest part of this paper, we will discuss these combination types and their dependence on the free parameters.

\begin{enumerate}
\item As we mentioned before, the principle coefficient $``s_{n-1}"$ determined the ring structure. Based on this, if only the preceeding coefficient $``s_{n-2}"$ is non zero, the high-order rogue waves have a double-ring structure. More explicitly, there are two rings with $2n-3$ single peaks in each and one $(n-4)$ basic mode in the middle (Figure 4). For example, (c) shows this type of combination for $|q_{6}|^2$. There are two outer ring structures with one 2-nd order basic mode in the center(Figure 1 (b)), and each individual ring is consisted of $9$ single peaks. When $n=4$, there is no fundamental pattern since $n-4=0$ (a).
\item $\emph {Triple-ring}$ structure is determined by the coefficient $s_{n-3}$.  When $s_{n-3}\neq0$, the order $n$ rogue wave has a triple ring structure with $n$ single peaks in the outer circular and an order $n-1$ double ring structure in the middle (Figure 5).
\item If $s_{1}$ and $s_{n-1}\neq0$, the phase space of high-order waves can be decomposed into two parts. The outer circular shell has $2n-1$ single peaks while the center is an order $n-2$ rogue wave with triangular structure (Figure 6).

\item If both $s_{n-1}$ and $s_{n-2}$ are non zero, then the order $n$ rogue wave is a composition of order $n$ and order $n-2$ ring structure such that the outer cell has $2n-1$ single peaks. And the inside is an order $(n-2)$ ring structures (Figure 7). Here we need to point out that although the shape is similar to double-ring structure described above, they are of different pattern in combination. In double-ring structure, both two outer circulars have $2n-3$ single peaks in each. While here, the outer cells have $2n-1$ and $2n-5$ single peaks in each respectively.

\item Other parameters also play an important role in determining the structure of high-order rogue waves. In other words, by choosing the parameter properly, we can get more complicated structures.
Here we give one more example. It is a composition of order $n$ ring structure and $n-2$ double-ring structure (Figure 8). Moreover, it is a complete decomposition in the ring structure. That is, each local maximum is of the same value and there is no fundamental pattern.
\end{enumerate}

\begin{remark}
There are several coincides between different patterns,
that is due to the special value of $n$. For example, when $n=4$,
then $n-3=1$, therefore, the triple-ring structure of the same shape
as ring structure is for triangular structure.
\end{remark}

Beside different structures, there are some properties that are preserved by all types of rogue wave solutions.
First, the total number of peaks in order $n$ solutions is $\frac{n(n+1)}{2}$. Secondly, the structures depend on every parameter. The principle coefficient $s_{n-1}$ leads to the $\emph {ring}$ structure. The first nontrivial coefficient $s_{1}$ leads to triangular solutions. They are very similar to the rational
solutions of the NLS equation apart from the finite tilt with regard to the axes. We find that the main effect of the third-order dispersion and a time-delay correction terms added to the NLS equation is to introduce a ``tilt" to the NLS equation solutions\cite{akhmediev3}, namely, the tilt is related to the finite value of parameter $\beta$, because if we set $\beta=0$ the tilt is zero. We can shorten the length of the ``ridge" by taking the value of $a$ properly, moreover, of interest it is to find that on the central line where $t=0$ the shape is independent of $\beta$ and $a$. That is, the adding the mKdV terms to the NLS equation do not have any impact in the figures of the rogue waves at the $t=0$ where the figures have its maximum. Consequently, the overall maximum value of the amplitude figure is $q_{j}(0,0)$ for any value of $\beta$, which is indeed the same as the NLS equation.

\section{Conclusions and discussions}
In this paper, we use the parameterized Darboux
transformation to construct the high-order rational solutions
for the Hirota equation. Based on the theoretical result, we classify the high-order rogue waves with respect to their intrinsic structure.
 We find out that there are three principle types, $\emph {basic mode}$, $\emph {ring}$ structure, and $\emph {triangular}$ structure.
 Moreover, the composition of the three principle structures are held by high-order rogue waves. As we mentioned in the very beginning,
 the Hirota equation is a generalized NLS equation. The ring structures obtained in this paper are similar to the ``atom" in Ref\cite{david}.
 This explains the generalization in terms of the solution. On the other hand, by changing the free parameters in the Darboux transformation we
 can get more complicated structures. It would be promising to ask whether there are some analogues between high-order rogue waves of different
 integrable equations? Also, it is essential to find more conserved properties for such kind of solutions. All of these will help us in a better understanding in the deep ocean wave.

 On the other hand, the most
remarkable feature of this equation is that it includes both the NLS
equation by restricting the parameter $\beta=0$ and the mKdV
equation by restricting the parameter $\alpha=0$. Namely, by taking
the different values of the parameters, we can also obtain
high-order rogue waves of the NLS equation and mKdV equation, with
regard to this point, we shall give the detailed describing in other
papers respectively.

\mbox{\vspace{1cm}}


{\bf Acknowledgments}

{\noindent \small This work is supported by the NSF of China under Grant No.11271210,  No.10971109 and K. C. Wong Magna Fund in Ningbo University. Jingsong He is also supported by  Natural Science Foundation of Ningbo under Grant No. 2011A610179. The authors appreciate the helpful suggestions of reviewers.}


\section{Appendix}
The formulas for $q_1,q_2$ are given under the condition
$s_0=0,\alpha=10,\beta=1,a=0,c=1$ and
$\alpha=5,\beta=1,s_0=0,a=3.88,c=1$ respectively.
\begin{eqnarray*}
q_1=e^{20it}\frac{3-1744\,{t}^{2}+48\,tx+160\,it-4\,{x}^{2}}{1+4\,{x}^{2}+1744\,{t}^{2}-48\,tx}\qquad\qquad\qquad \\
\end{eqnarray*}
\vspace*{-30pt}

\begin{equation*}
\begin{aligned}
q_2=&{{\rm e}^{i \left(  3.88\,x- 30.140928\,t \right) }}\frac{\Delta_1}{\Delta_2}\\
\Delta_1=&45+ 351839417.0\,{t}^{6}-
252396.4448\,{t}^{2}{x}^{2}+33987.4508\,{t} ^{2}{x}^{4}\\
&+ 139.4688\,{x}^{5}t+ 4344315.56\,{t}^{5}x+
49254.311\,{x}^{3}{t}^{3}\\
&+1326.7968\,{x}^{3}t+ 5998449.26\,{t}^{4}{x}^{2}-
 996552.958\,x{t}^{3}\\
 &+ 3325.2480\,tx+ 1307537.37\,i{t}^{4}x- 127944.917\,i{t}^{2}x\\
 &- 4780.80\,it+ 7408.583\,i{t}^{2}{x}^{3}+ 1802722.57\,i{t} ^{3}{x}^{2}\\
 &+144\,is_{{1}}-180\,{x}^{2}- 72088.2566\,{t}^{2}+ 404211.06\,i{t}^{3}\\
 &+158844096.7\,i{t}^{5}- 16750593.96\,{t}^{4}+1912.32\,ts_{{1}}\\
 &-144\,{x}^{4}+5556.437\,{t}^{2}xs_{{1}}+ 7649.28\,t{x}^{2}s_{{1}}+144\,{s_{{1}}}^{2}\\
&+ 418.4064\,itxs_{{1}}+ 5099.52\,it{x}^{4}- 7649.28\,it{x}^{2}+64\,{x}^{6}\\
&-448662.55\,{t}^{3}s_{{1}}-101506.4548\,i{t}^{2}s_{{1}}+576\,is_{{1}}
{x}^{2}
\end{aligned}
\end{equation*}

\begin{equation*}
\begin{aligned}
\Delta_2=&9+ 351839351.0\,{t}^{6}-
49079.6020\,{t}^{2}{x}^{2}+33987.4508\,{t}^{ 2}{x}^{4}\\
&- 5736.96\,ts_{{1}}- 448662.552\,{t}^{3}s_{{1}}+ 5998449.36\,
{t}^{4}{x}^{2}\\
&+ 4344308.56\,{t}^{5}x+ 49254.311\,{x}^{3}{t}^{3}+
 1605.7344\,{x}^{3}t\\
 &+ 139.4688\,{x}^{5}t- 848937.199\,x{t}^{3}-
 1073.5488\,tx\\
 &+108\,{x}^{2}+ 78649.7667\,{t}^{2}+48\,{x}^{4}+
 13134403.21\,{t}^{4}\\
 &+ 5556.437\,{t}^{2}xs_{{1}}+ 7649.28\,t{x}^{2}s_{
{1}}+64\,{x}^{6}+144\,{s_{{1}}}^{2}
\end{aligned}
\end{equation*}

On the other hand, we also give the formulas for $q_2$ under the
condition $\alpha=5,\beta=1,s_1=0,a=3.88,c=1$.

\begin{equation*}
\begin{aligned}
q_2=&{{\rm e}^{i (  3.88\,x- 30.140928\,t ) }}\frac{\Delta_3}{\Delta_4}\\
\Delta_3=&45+
5099.52\,it{x}^{4}-7649.28\,it{x}^{2}+64\,{x}^{6}-144\,{x}^{4}\\
&- 180\,{x}^{2}+ 351839311.0\,{t}^{6}-
16750593.36\,{t}^{4}\\
&-72088.2566\, {t}^{2}+ 404211.06\,i{t}^{3} +
158844096.7\,i{t}^{5}\\
&+ 5998449.06\,{t}^{ 4}{x}^{2}+ 33987.4508\,{t}^{2}{x}^{4}-
252396.4448\,{t}^{2}{x}^{2}\\
&- 996552.948\,{t}^{3}x+ 49254.311\,{t}^{3}{x}^{3}+ 3325.2480\,tx\\
&+ 139.4688\,t{x}^{5}+ 4344348.56\,{t}^{5}x+ 1326.7968\,t{x}^{3}\\
&- 4780.80\,it+ 7408.583\,i{t}^{2}{x}^{3}+ 1307537.37\,i{t}^{4}x\\
&- 127944.917\,i{t}^{2}x+ 1802722.57\,i{t}^{3}{x}^{2}+
49254.311\,{s_{{0 }}}^{3}{t}^{3}\\
&- 996552.928\,s_{{0}}{t}^{3}+ 1326.7968\,{s_{{0}}}^{3}t+
 139.4688\,{s_{{0}}}^{5}t\\
 &+ 5998446.04\,{t}^{4}{s_{{0}}}^{2}+
 33987.4508\,{s_{{0}}}^{4}{t}^{2}+ 3325.2480\,s_{{0}}t\\
 &- 252396.4448\,{
s_{{0}}}^{2}{t}^{2}-576\,x{s_{{0}}}^{3}+384\,{x}^{5}s_{{0}}+960\,{x}^{
4}{s_{{0}}}^{2}\\
&+1280\,{x}^{3}{s_{{0}}}^{3}-864\,{x}^{2}{s_{{0}}}^{2}+
384\,{s_{{0}}}^{5}x+960\,{s_{{0}}}^{4}{x}^{2}\\
&+ 4344312.56\,{t}^{5}s_{{0}}-
15298.56\,is_{{0}}tx+ 30597.12 \,i{s_{{0}}}^{2}t{x}^{2}\\
&+ 20398.08\,is_{{0}}t{x}^{3}+ 20398.08\,i{s_{{0 }}}^{3}tx+
22225.748\,i{s_{{0}}}^{2}{t}^{2}x\\
&+ 22225.748\,i{t}^{2}{x}^{ 2}s_{{0}}+
3605445.03\,i{t}^{3}xs_{{0}}-180\,{s_{{0}}}^{2}-144\,{s_{{0
}}}^{4}\\
&- 504792.8896\,{t}^{2}s_{{0}}x+ 11996906.61\, {t}^{4}s_{{0}}x+
135949.802\,{t}^{2}{x}^{3}s_{{0}}\\
&+ 203924.713\,{t}^{2 }{x}^{2}{s_{{0}}}^{2}+
135949.802\,{s_{{0}}}^{3}{t}^{2}x+ 697.3440\,{s _{{0}}}^{4}tx\\
&+ 1394.6880\,{s_{{0}}}^{3}t{x}^{2}+ 697.3440\,{x}^{4}s_{{0 }}t+
1394.6880\,{x}^{3}{s_{{0}}}^{2}t\\
&+ 3980.3904\,{x}^{2}s_{{0}}t+
 3980.3904\,x{s_{{0}}}^{2}t+ 147763.18\,{t}^{3}{s_{{0}}}^{2}x\\
 &+
 147763.15\,{t}^{3}s_{{0}}{x}^{2}+ 7408.583\,i{s_{{0}}}^{3}{t}^{2}+
 1307537.37\,is_{{0}}{t}^{4}\\
 &- 127944.917\,is_{{0}}{t}^{2}+ 1802722.57
\,i{s_{{0}}}^{2}{t}^{3}-360 \,s_{{0}}x\\
&+64\,{s_{{0}}}^{6}+ 5099.52\,i{s_{{0}}}^{4}t- 7649.28\,i{s_{{0}}}
^{2}t-576\,{x}^{3}s_{{0}}
\end{aligned}
\end{equation*}

\begin{equation*}
\begin{aligned}
\Delta_4=&9+64\,{x}^{6}+48\,{x}^{4}+108\,{x}^{2}+
351839241.0\,{t}^{6}\\
&+13134403.54\,{t}^{4}+ 78649.7667\,{t}^{2}+
5998445.94\,{t}^{4}{x}^{2}\\
& + 33987.4508\,{t}^{2}{x}^{4}- 49079.6020\,{t}^{2}{x}^{2}-
848937.199\, {t}^{3}x\\
&+ 49254.351\,{t}^{3}{x}^{3}- 1073.5488\,tx+ 139.4688\,t{x}^{5}\\
& + 4344312.56\,{t}^{5}x+ 1605.7344\,t{x}^{3}+
49254.311\,{s_{{0}}}^{3}{ t}^{3}\\
&- 848937.199\,s_{{0}}{t}^{3}+ 1605.7344\,{s_{{0}}}^{3}t+
 139.4688\,{s_{{0}}}^{5}t\\
 &+ 5998445.94\,{t}^{4}{s_{{0}}}^{2}+
 33987.4508\,{s_{{0}}}^{4}{t}^{2}- 1073.5488\,s_{{0}}t\\
 &- 49079.6018\,{s
_{{0}}}^{2}{t}^{2}+192\,x{s_{{0}}}^{3}+384\,{x}^{5}s_{{0}}+960\,{x}^{4
}{s_{{0}}}^{2}\\
&+1280\,{x}^{3}{s_{{0}}}^{3}+288\,{x}^{2}{s_{{0}}}^{2}+
384\,{s_{{0}}}^{5}x+960\,{s_{{0}}}^{4}{x}^{2}+192\,{x}^{3}s_{{0}}\\
\end{aligned}
\end{equation*}

\begin{equation*}
\begin{aligned}
&+216 \,s_{{0}}x+
4344308.56\,{t}^{5}s_{{0}}+108\,{s_{{0}}}^{2}+48\,{s_{{0}}
}^{4}+64\,{s_{{0}}}^{6}\\
&- 98159.2036\,{t}^{2}s_{{0}}x+ 11996906.81\,{t} ^{4}s_{{0}}x+
135949.802\,{t}^{2}{x}^{3}s_{{0}}\\
&+ 203924.713\,{t}^{2}{x }^{2}{s_{{0}}}^{2}+
135949.802\,{s_{{0}}}^{3}{t}^{2}x+ 697.3440\,{s_{{0 }}}^{4}tx\\
&+ 1394.6880\,{s_{{0}}}^{3}t{x}^{2}+ 697.3440\,{x}^{4}s_{{0}}t +
1394.6880\,{x}^{3}{s_{{0}}}^{2}t\\
&+ 4817.2032\,{x}^{2}s_{{0}}t+
 4817.2032\,x{s_{{0}}}^{2}t+ 147763.15\,{t}^{3}{s_{{0}}}^{2}x\\
 &+
 147763.15\,{t}^{3}s_{{0}}{x}^{2}
\end{aligned}
\end{equation*}

\begin{figure}[!ht]
\centering
\renewcommand{\figurename}{Fig.}
\raisebox{0.85in}{($|q_{1}|^2$)}\includegraphics[height=5cm,width=5cm]{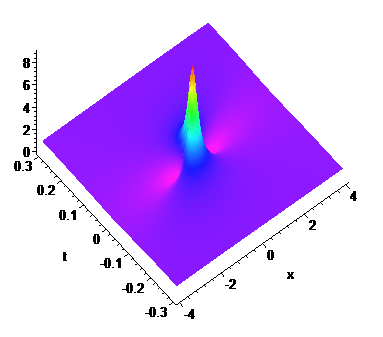}%
\hspace{1cm}
\raisebox{0.85in}{($|q_{2}|^2$)} \includegraphics[height=5cm,width=5cm]{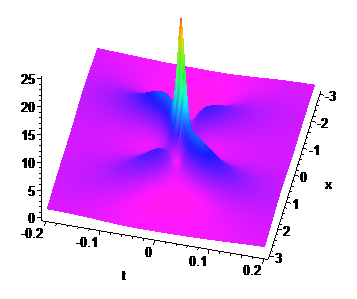}\\%
 \renewcommand{\figurename}{Fig.}
\raisebox{0.85in}{($|q_{3}|^2$)}\includegraphics[height=5cm,width=5cm]{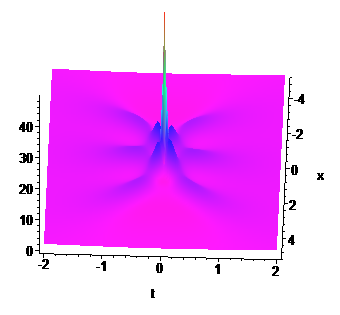}%
\hspace{1cm}
\raisebox{0.85in}{($|q_{6}|^2$)}\includegraphics[height=5cm,width=5cm]{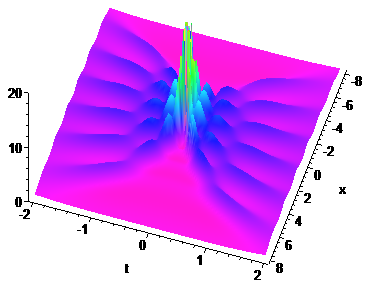}%
 \begin{center}
\parbox{15.5cm}{\small{\bf Figure 1.}\ \
\ Higher order rogue waves with zero shifts ($\emph{fundamental
pattern}$), i.e., (a) $n=1, \alpha=10,\beta=1,a=0,c=1,s_0=0$. (b)
$n=2, \alpha=5,\beta=1,a=3.88,c=1,s_0=0,s_1=0$. (c) $n=3,
\alpha=1,\beta=\frac{1}{50},a=0,c=1,s_0=0,s_1=0,s_2=0$.
 (d) $n=6,\alpha=1,\beta=\frac{1}{50},a=0,c=1,s_0=0,s_1=0,s_2=0,s_3=0,s_4=0,s_5=0$. Here in the figure of $|q_{6}|^2$, to show the lower peaks clearly, the whole amplitude of the maximum (central peak) is cut. The maximum is $169$ (that is $(2n+1)^2$).}
\end{center}
\end{figure}
\textheight 24cm
\begin{figure}[!ht]
 \centering
\renewcommand{\figurename}{Fig.}
\raisebox{0.85in}{($|q_{3}|^2$)}\includegraphics[height=5cm,width=5cm]{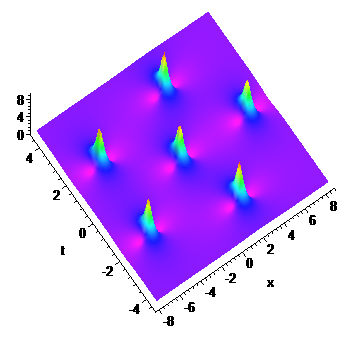}%
\hspace{1cm}
\raisebox{0.85in}{($|q_{3}|^2$)}\includegraphics[height=5cm,width=5cm]{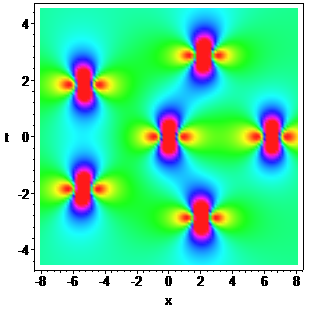}\\%
\raisebox{0.85in}{($|q_{4}|^2$)}\includegraphics[height=5cm,width=5cm]{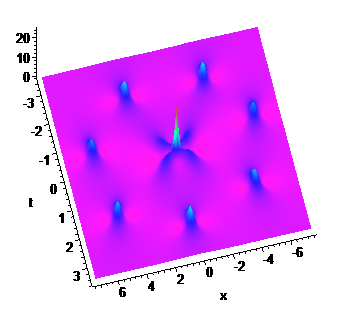}%
\hspace{1cm}
\raisebox{0.85in}{($|q_{4}|^2$)} \includegraphics[height=5cm,width=5cm]{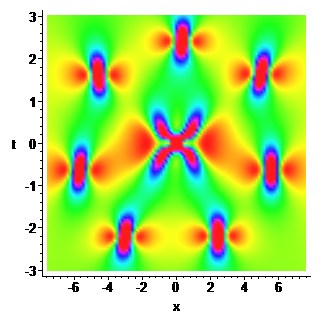}\\%
\raisebox{0.85in}{($|q_{5}|^2$)} \includegraphics[height=5cm,width=5cm]{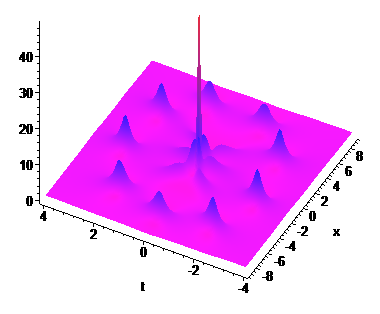}%
\hspace{1cm}
\raisebox{0.85in}{($|q_{5}|^2$)} \includegraphics[height=5cm,width=5cm]{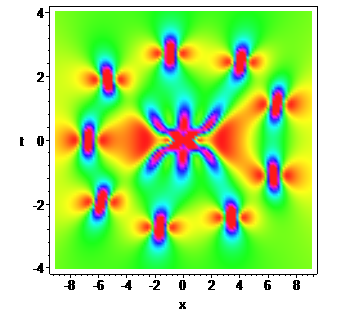}\\%
\raisebox{0.85in}{($|q_{6}|^2$)} \includegraphics[height=5cm,width=5cm]{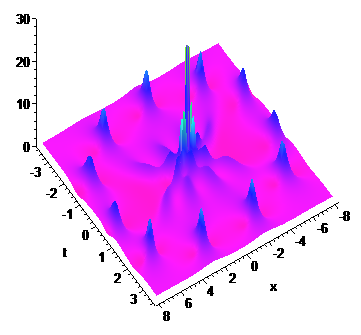}%
\hspace{1cm}
\raisebox{0.85in}{($|q_{6}|^2$)} \includegraphics[height=5cm,width=5cm]{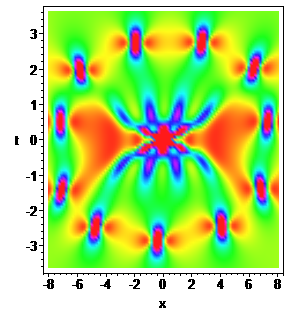}%
 \begin{center}
\parbox{15.5cm}{\small{\bf Figure 2.}\ \ Single ring structure of $|q_n|^2$ (left) and
 the corresponding density plot (right) under the condition (a)
$n=3, \alpha=1, \beta=\frac{1}{50}, a=-\frac{1}{20}, c=1, s_0=0,
s_1=0, s_2=1000$. The trajectory is a pentagon; (b) $n=4,
\alpha=1,\beta=\frac{1}{50}, a=0, c=1, s_0=0, s_1=0, s_2=0,
s_3=1000$; (c) $n=5, \alpha=1, \beta=\frac{1}{50}, a=0, c=1, s_0=0,
s_1=0, s_2=0, s_3=0, s_4=4500$; (d) $n=6, \alpha=1,
\beta=\frac{1}{50}, a=0, c=1, s_0=0, s_1=0, s_2=0, s_3=0, s_4=0,
s_5=10000$.  Here in the figure of $|q_{6}|^2$, to show the lower
peaks clearly, the whole amplitude of the maximum (central peak) is
cut. The maximum of the inner order 4 rogue wave is $81$.}
\end{center}
\end{figure}
\begin{figure}[!ht]
\centering
\renewcommand{\figurename}{Fig.}
\raisebox{0.85in}{($|q_{2}|^2$)}\includegraphics[height=5cm,width=5cm]{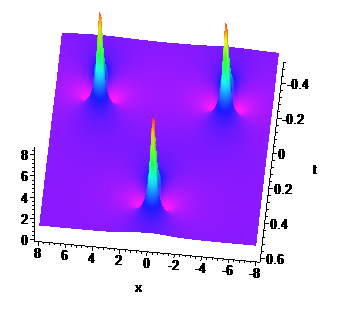}%
\hspace{1cm}
 \raisebox{0.85in}{($|q_{2}|^2$)}\includegraphics[height=5cm,width=5cm]{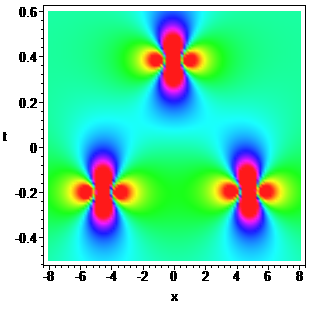}\\%
 \raisebox{0.85in}{($|q_{3}|^2$)}\includegraphics[height=5cm,width=5cm]{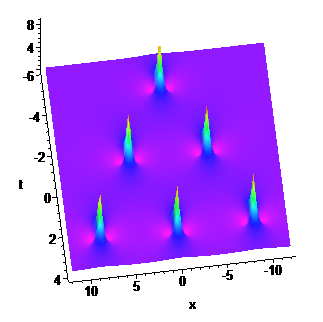}%
\hspace{1cm}
 \raisebox{0.85in}{($|q_{3}|^2$)}\includegraphics[height=5cm,width=5cm]{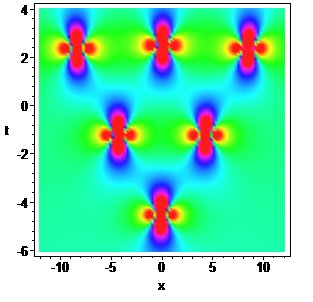}\\%
 \raisebox{0.85in}{($|q_{4}|^2$)}\includegraphics[height=5cm,width=5cm]{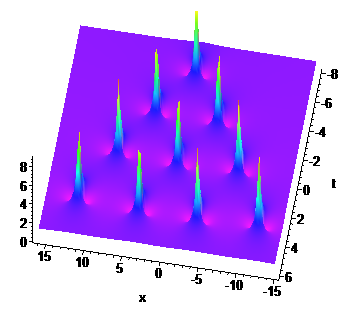}%
\hspace{1cm}
\raisebox{0.85in}{($|q_{4}|^2$)} \includegraphics[height=5cm,width=5cm]{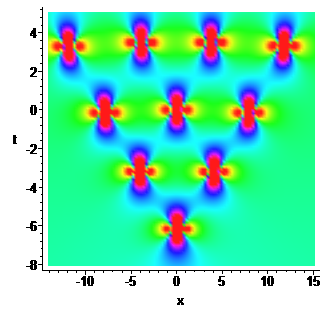}\\%
\raisebox{0.85in}{($|q_{6}|^2$)} \includegraphics[height=5cm,width=5cm]{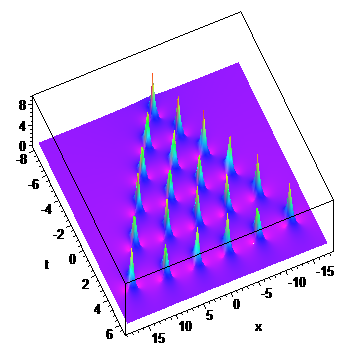}%
\hspace{1cm}
\raisebox{0.85in}{($|q_{6}|^2$)} \includegraphics[height=5cm,width=5cm]{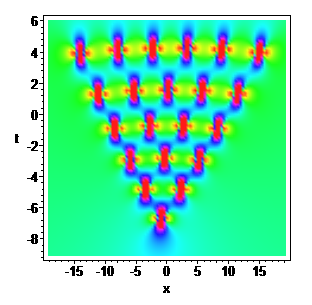}%
 \begin{center}
\parbox{15.5cm}{\small{\bf Figure 3.}\ \ Triangular structure of $|q_n|^2$
 (a) $n=2, \alpha=5, \beta=1, a=3.88, c=1, s_0=0, s_1=100$;
(b) $n=3,
\alpha=1,\beta=\frac{1}{50},a=-\frac{1}{20},c=1,s_0=0,s_1=100,s_2=0$;
(c) $n=4,
\alpha=1,\beta=\frac{1}{50},a=-\frac{1}{15},c=1,s_0=0,s_1=100,s_2=0,s_3=0$;
(d) $n=6,
\alpha=1,\beta=\frac{1}{50},a=0,c=1,s_0=0,s_1=50,s_2=0,s_3=0,s_4=0,s_5=0$.}
\end{center}
\end{figure}
\begin{figure}[!ht]
\centering
\renewcommand{\figurename}{Fig.}
 \raisebox{0.85in}{($|q_{4}|^2$)}\includegraphics[height=5cm,width=5cm]{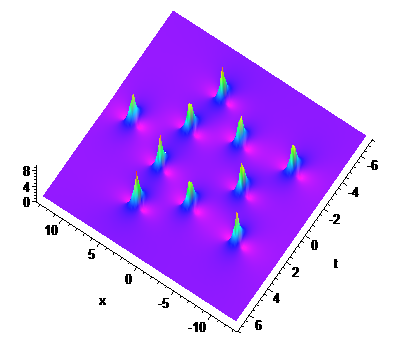}%
\hspace{1cm}
 \raisebox{0.85in}{($|q_{4}|^2$)} \includegraphics[height=5cm,width=5cm]{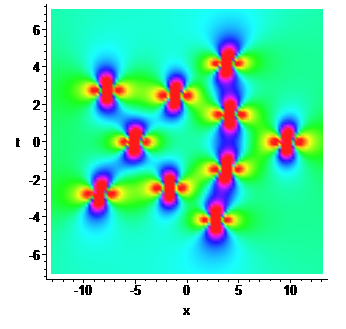}\\%
 \raisebox{0.85in}{($|q_{5}|^2$)} \includegraphics[height=5cm,width=5cm]{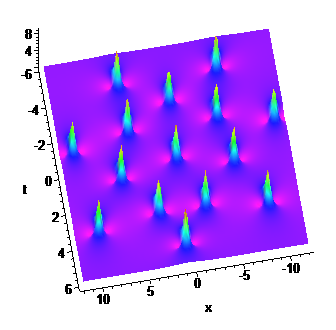}%
\hspace{1cm}
 \raisebox{0.85in}{($|q_{5}|^2$)} \includegraphics[height=5cm,width=5cm]{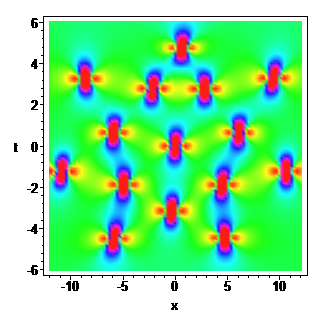}\\%
 \raisebox{0.85in}{($|q_{6}|^2$)} \includegraphics[height=5cm,width=5cm]{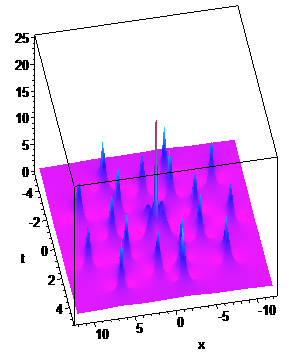}%
\hspace{1cm}
 \raisebox{0.85in}{($|q_{6}|^2$)} \includegraphics[height=5cm,width=5cm]{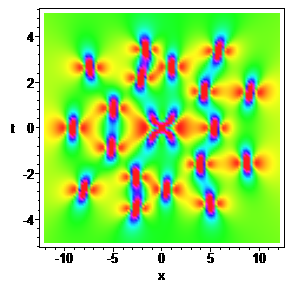}%

 \begin{center}
\parbox{15.5cm}{\small{\bf Figure 4.}\ \ The decomposition of $|q_n|^2$ into double-ring structure (a)
$n=4,
\alpha=1,\beta=\frac{1}{50},a=0,c=1,s_0=0,s_1=0,s_2=1000,s_3=0$; (b)
$n=5,
\alpha=1,\beta=\frac{1}{50},a=0,c=1,s_0=0,s_1=0,s_2=0,s_3=10000,s_4=0$;
(c) $n=6,
\alpha=1,\beta=\frac{1}{50},a=0,c=1,s_0=0,s_1=0,s_2=0,s_3=0,s_4=6000,s_5=0$.
}
\end{center}
\end{figure}
\begin{figure}[!ht]
\centering
\renewcommand{\figurename}{Fig.}
\raisebox{0.85in}{($|q_{5}|^2$)}\includegraphics[height=5cm,width=5cm]{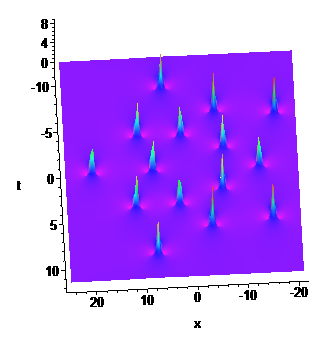}%
\hspace{1cm}
\raisebox{0.85in}{($|q_{5}|^2$)} \includegraphics[height=5cm,width=5cm]{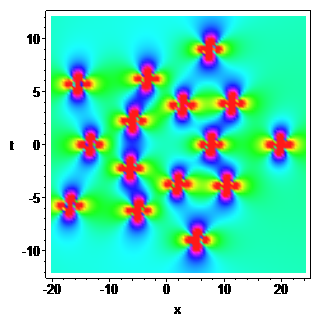}\\%
\raisebox{0.85in}{($|q_{6}|^2$)} \includegraphics[height=5cm,width=5cm]{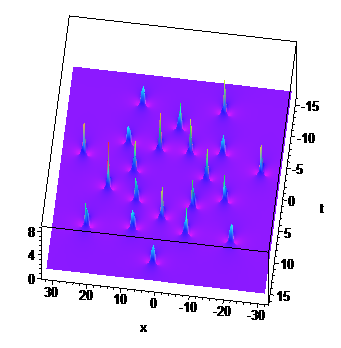}%
\hspace{1cm}
\raisebox{0.85in}{($|q_{6}|^2$)} \includegraphics[height=5cm,width=5cm]{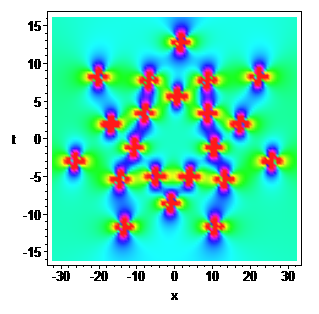}%
 \begin{center}
\parbox{15.5cm}{\small{\bf Figure 5.}\ \ Triple-ring structure of $|q_n|^2$ (a)
$n=5,
\alpha=1,\beta=\frac{1}{50},a=0,c=1,s_0=0,s_1=0,s_2=8000,s_3=0,s_4=0$;
(b) $n=6,
\alpha=1,\beta=\frac{1}{50},a=0,c=1,s_0=0,s_1=0,s_2=0,s_3=1000000,s_4=0,s_5=0$.
}
\end{center}
\end{figure}

\begin{figure}[!ht]
\centering
\renewcommand{\figurename}{Fig.}
\raisebox{0.85in}{($|q_{4}|^2$)}\includegraphics[height=5cm,width=5cm]{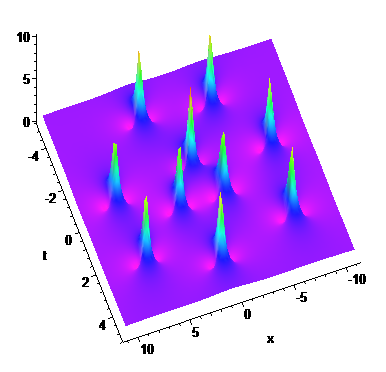}%
\hspace{1cm}
\raisebox{0.85in}{($|q_{4}|^2$)}\includegraphics[height=5cm,width=5cm]{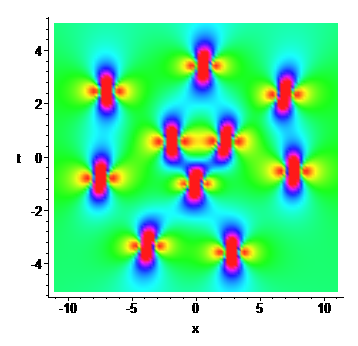}\\%
\raisebox{0.85in}{($|q_{5}|^2$)} \includegraphics[height=5cm,width=5cm]{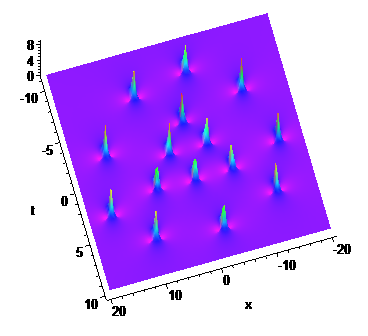}%
\hspace{1cm}
\raisebox{0.85in}{($|q_{5}|^2$)} \includegraphics[height=5cm,width=5cm]{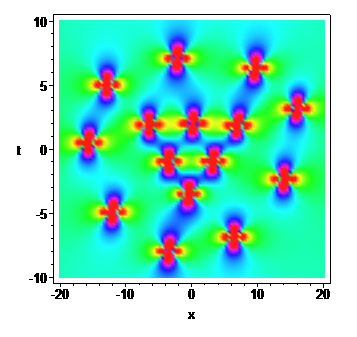}\\%
 \raisebox{0.85in}{($|q_{6}|^2$)}\includegraphics[height=5cm,width=5cm]{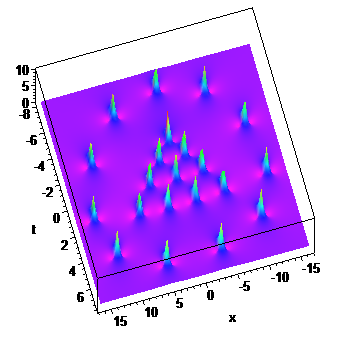}%
\hspace{1cm}
 \raisebox{0.85in}{($|q_{6}|^2$)}\includegraphics[height=5cm,width=5cm]{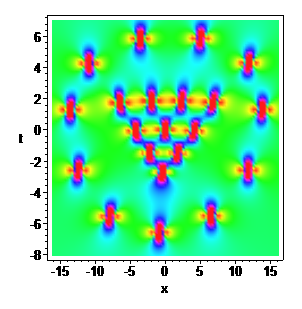}%

 \begin{center}
\parbox{15.5cm}{\small{\bf Figure 6.}\ \ Ring plus triangular decomposition of $|q_n|^2$ (a)
$n=4,
\alpha=1,\beta=\frac{1}{50},a=0,c=1,s_0=0,s_1=10,s_2=0,s_3=10000$;
(b) $n=5,
\alpha=1,\beta=\frac{1}{50},a=0,c=1,s_0=0,s_1=50,s_2=0,s_3=0,s_4=10000000$;
(c) $n=6,
\alpha=1,\beta=\frac{1}{50},a=0,c=1,s_0=0,s_1=15,s_2=0,s_3=0,s_4=0,s_5=10000000$.}
\end{center}
\end{figure}
\begin{figure}[!ht]
\centering
\renewcommand{\figurename}{Fig.}
\raisebox{0.85in}{($|q_{5}|^2$)}\includegraphics[height=5cm,width=5cm]{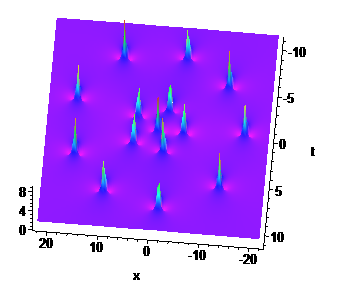}%
\hspace{1cm}
\raisebox{0.85in}{($|q_{5}|^2$)} \includegraphics[height=5cm,width=5cm]{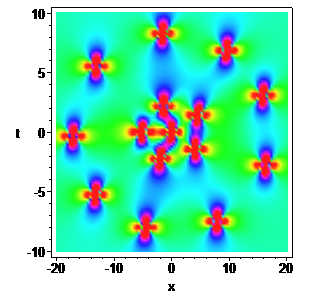}\\%
\raisebox{0.85in}{($|q_{6}|^2$)}\includegraphics[height=5cm,width=5cm]{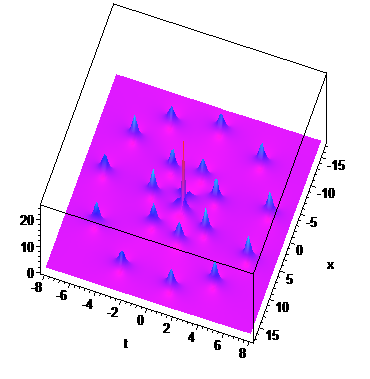}%
\hspace{1cm}
\raisebox{0.85in}{($|q_{6}|^2$)} \includegraphics[height=5cm,width=5cm]{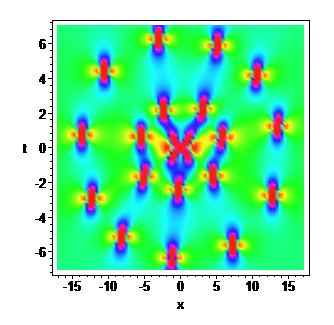}%
 \begin{center}
\parbox{15.5cm}{\small{\bf Figure 7.}\ \ Fundamental pattern plus ring structure decomposition (a)
$n=5,
\alpha=1,\beta=\frac{1}{50},a=0,c=1,s_0=0,s_1=0,s_2=0,s_3=80000,s_4=20000000$;
(b) $n=6,
\alpha=1,\beta=\frac{1}{50},a=0,c=1,s_0=0,s_1=0,s_2=0,s_3=0,s_4=100000,s_5=10000000$.}
\end{center}
\end{figure}
\begin{figure}[!ht]
\centering
\renewcommand{\figurename}{Fig.}
\raisebox{0.85in}{($|q_{6}|^2$)}\includegraphics[height=5cm,width=5cm]{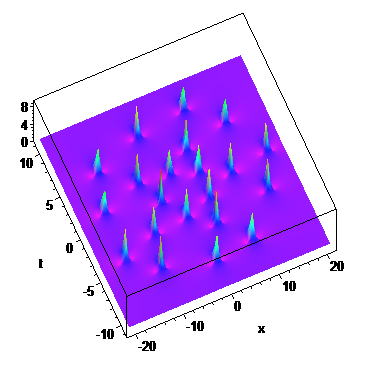}%
\hspace{1cm}
\raisebox{0.85in}{($|q_{6}|^2$)} \includegraphics[height=5cm,width=5cm]{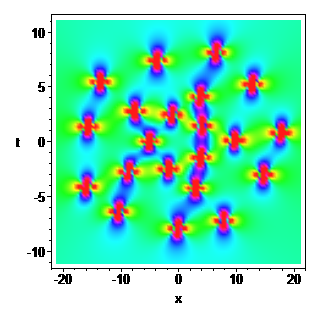}%
 \begin{center}
\parbox{15.5cm}{\small{\bf Figure 8.}\ \ Complete decomposition of $|q_6|^2$ into rings of single peaks
$\alpha=1,\beta=\frac{1}{50},a=0,c=1,s_0=0,s_1=0,s_2=1000,s_3=0,s_4=0,s_5=100000000$.
}
\end{center}
\end{figure}

\end{document}